\documentclass[preprint]{article}
\usepackage{graphicx}
\usepackage{dcolumn}
\usepackage{bm}
\usepackage{amsmath}
\usepackage{mathrsfs}
\usepackage{amssymb}
\usepackage{multirow}
\usepackage{caption}
\usepackage{epstopdf}
\usepackage{hyperref}
\usepackage[noadjust]{cite}
\usepackage{hyperref}
\hypersetup{colorlinks=true, linkcolor=red, citecolor=blue}
\usepackage[table]{xcolor}
\begin{document}
{\setlength{\oddsidemargin}{1.2in}
\setlength{\evensidemargin}{1.2in} } \baselineskip 0.55cm
\begin{center}
{\LARGE {\bf Gravastars with Kuchowicz Metric Potential in $f(R, \Sigma, T)$ Gravity}}
\end{center}
\date{\today}
\begin{center}
  Bharat Singh, S. Surendra Singh \\
   Department of Mathematics, National Institute of Technology, Manipur,\\ Imphal-795004,India\\
   Email:{ bharatsinghiam@gmail.com, ssuren.mu@gmail.com}\\
 \end{center}

\begin{center}
\textbf{Abstract}
\end{center}
This manuscript explores the gravastar model in the $f(R,\Sigma,T)$ gravity framework, with the help of Kuchowicz metric funcition, offering an alternative to black holes. A gravastar has three regions: interior, intermediate shell, and exterior. The interior region has pressure equal to negative density, generating a repulsive force across the thin shell. The intermediate shell contains ultra-relativistic plasma fluids, with pressure proportional to density, balancing the interior's repulsive force. The exterior region is a vacuum, described by a generalized Schwarzschild solution. Our specifications yield precise, singularity-free gravaster solutions with physically valid features in the $f(R,\Sigma,T)$ gravity framework, exploring strong gravity and anti-gravity aspects. The gravitational Lagrangian is based on an arbitrary function of torsion scalar $\Sigma$ and  trace of the energy-momentum tensor $T$. Our $f(R,\Sigma,T)$ gravity analysis explores gravastars inner workings, revealing insights into gravity, strong gravity, and antigravity forces due to torsion effects. We examine shell properties like length, energy, entropy, and discussed junction conditions. Key findings include constant interior density and pressure, denser shell fluid at the outer boundary, and increasing shell length. These results illuminate gravastar behavior and fundamental gravitational principles.\\


\section{Introduction}

\hspace{0.5cm}Black holes, a fascinating object in Einsteins's GR, are well-studied by astrophysicists. Black holes from collaspsing stars, with their type determined by the star's mass, marking the final stage of a star's life. Black holes have intense gravitational pull due to their immense density and compact volume, trapping everything, including light. Black hole astrophysics emerged in the 1960s following key discoveries.
Black hole research gained momentum with the 1970s discovery of Cygnus X-1. Now, numerous stellar-mass and supermassive black hole candidates are known. Classical General Relativity fullly describes completely \cite{Misner1973,Shapiro1983,Wald1984,Carroll1996,Townsend1997,Narayan2005}. However, despite two major issues: (1) event horizons and (2) singularities at their centers. To address black hole issues, Mazur and Mottola \cite{Mazur2001} (2001) introduced Gravastars, extremely compact objects, as an alternative. Mazur and Mottola explored Bose-Einstein condensation in gravitational systems, where particles condense at extremely low temperatures. Using this concept, they proposed Gravastars: hypothetical, cold, compact, and dark object that avoid the limitations and issues of classical black holes. Gravastars emerged as a promising alternative  to classical black hole sparking significant interest and research in the field. At low temperatures, a phase transition creates a repulsive de sitter core, preventing event horizon formation and singularity \cite{Gliner,Chapline}. This transition likely occurs near the limit where $\frac{2m(r)}{r}$ = 1, making it nearly indistinguishable from a black hole for outside observers. In Mazur and Mottola's gravastar model, quantum vacuum fluctuations significantly influence collapse dynamics. The Mazur-Mottola gravastar model is thermodynamically stable and avoids the black hole information paradox. The gravastar model consists of:
(I) An isotropic de Sitter vacuum interior.
(II) A Schwarzschild geometry exterior.
The gravaster model has three regions with distinct equations of state (EoS):
(I). Interior (de sitter): $p=-\rho$\\ (II). Shell (stiff matter): $p=\rho$\\ (III). Exterior (Schwarzschild): $p=\rho=0$ (vacuum)\\ These EoS define the properties of each region.  
So, $R_{1}$ and $R_{2}$ are the boundaries of the thin shell: $R_{1}$: Inner radius (interface with de sitter interior) $R_{2}$ Outer radius (interface with Schwarzschild exterior) The thin shell lies between $R_{1}$ and $R_{2}$. The stiff matter shell with thickness $R_{2} - R_{1}=\epsilon$ provides the necessary inward pressure to counterbalance the repulsive force from the de Sitter interior, ensuring stability.  

The dark energy (or vacuum energy ) \cite{Perlmutter,Riess} in the gravastar's interior generates negative pressure, exerting a repulsive force outward from the center. This outward pressure is counterbalanced by the inward pressure from the stiff matter shell, maintaining stability. The shell's positive matter density creates an inward gravitational pull, balancing the outward repulsive force from the interior's dark energy. This balance maintains the gravastar's stability. The interior's equation of state (EoS) leads to degenerate energy states, characterized as a degenerate vacuum, false vacuum, or $\rho$-vacuum in literature \cite{Davies,Kaiser}. This suggests a metastable  state with potentially interesting  implications.
Gliner introduced this type of EoS to study the energy-momentum tensor's algebraic properties, laying groundwork for later research on gravastars and related concepts. The same EoS is relevant in the Casimir Effect \cite{Casimir}, where it helps to the gravitational impact of zero-point energies of particles and electromagnetic fields. This highlights the EoS's broader applicability in physics. References \cite{Dymnikova1992,Dymnikova1998,Dymnikova2001,Dymnikova2003} explore spherically symmetric systems with de Sitter asymptotics, while the stiff fluid model in the shell is reminiscent of Zel’dovich's  work on cold baryonic universe models \cite{Zel’dovich}. The stiff fluid model has seen diverse applications:
(I) Cosmology \cite{Madsenet,Carr,Chakraborty}
(II) Astrophysics \cite{Linares,Braje,Buchert}
Its versatility makes it useful across these fields.
The gravastar's exterior is a vacuum spacetime, described by the Schwarzschild solution with equation of state $p = \rho = 0$, indicating no matter or radiation. \\
Mazur and Mottola's proposal sparked significant research on gravastars, exploring their properties and implications in gravity and cosmology. Mazur and Mottola demonstrated thermodynamic stability in their five-layer gravastar model, a key finding in gravastar research. Visser and Wiltshire \cite{Visser} simplified the Mazur-Mottola model to a three-layer gravastar and demonstrated its dynamical stability against spherically symmetric perturbations. Carter \cite{Carter} generalized the stability analysis of gravastars to include models with various exterior configurations, expanding on previous work. Researchers have explored gravastar stability through:
1. Axial perturbations \cite{DeBenedictis},
2. Linearized stability in non-commutative geometry \cite{Lobo}.
These studies provide insights into gravastar stability under different conditions.\\
Researchers have explored gravastars in modified gravity theories, specifically $f(R, T$ gravity, investigating  electromagnetic field effects on isotropic spherical gravastar models \cite{Yousaf2019}. 

This shows the interest in gravastars beyond standard gravity frameworks.
Yousaf et al. and Das et al. proposed gravastar models in $f(T)$ gravity as alternatives to black holes, building on Mazur and Mottola's conjecture. There is extensive research on gravastars in the literature, with various works exploring mathematical and physical aspects in different gravity contexts, as seen in references \cite{Majeed2022}. Many researchers have explored gravastars within Einstein's general relativity framework, as seen in references \cite{Turimov2009,Garattini2013,Bhar2014}. GR has been instrumental in understanding the universe, but observational evidence suggests that the universe is accelerating and DM exists \cite{Padmanabhan2003,Clifton2012}, pointing to areas for further exploration. 
This paper explores gravastars in a modified gravity theory incorporating torsion, aiming to address the dark energy issue within the gravastar's inner region. Torsion can be dynamic, contributing to gravitation alongside curvature, and theories incorporating torsion can be consistent with the Weak Equivalence Principle (WEP). This article reviews the geometry of Absolute Parallelism (AP) space to explore the role of torsion tensor and scalar in gravastar evolution. This research will examine gravastar behavior within the framework of $f(R, \Sigma, T)$ gravity and antigravity theories.

\section{An summary of the AP geometry in brief}
\hspace{0.5cm}We go into AP geometry in brief in this part, which suggests an interaction known as the ``antigravity interaction." We'll give a quick overview of the evolution of this geometry before talking about how antigravity affects space-time. The vector $\lambda^\Upsilon_{i} ( 1, 2, 3, 4 = i)$ indicates the vector number and defines the entire building of the conventional AP in four dimensions. Furthermore, the coordinate component are shown as $(1,2,3,4=\Upsilon)$. The covariant vector of $\lambda_i^\Upsilon$ can be found using \cite{Mikhail1977,Mikhail1962,Wanas1997}.
\begin{equation}\label{1a}
\lambda^\Upsilon_{i} \ \lambda_i^\mu= \ \delta^\Upsilon_\mu, \ 
\lambda_\mu^i \ \lambda_j^\mu\ = \ \delta_{ij}.
\end{equation}
At any stage in AP geometry, one can delineate the line element as
\begin{equation}\label{2a}
 ds^2 = g_{\mu\nu}  dx^{\mu} dx^{\nu}
\end{equation}
including the definitions

\begin{equation}\label{3a}
g_{\mu\nu} = \eta_{ij}\,\lambda_{i}{}_{\mu}\lambda_{j}{}_{\nu}, \\
g^{\mu\nu} = \eta_{ij}\,\lambda_{i}{}^{\mu}\lambda_{j}{}^{\nu}, \\
g^{\mu\nu} g_{\nu\Upsilon} = \delta^{\mu}{}_{\Upsilon},
\end{equation}
where $\eta_{ij} = \mathrm{diag}(1, -1, -1, -1)$ is the Minkowski metric. The following Levi-Civita link is used in GR:  

\begin{equation}\label{4a}
\Gamma^{\Upsilon}_{\mu\nu} 
= \frac{1}{2} g^{\Upsilon \epsilon} 
\left( g_{\mu \epsilon,\nu} + g_{\nu \epsilon,\mu} - g_{\mu \nu,\epsilon} \right)
\end{equation}

As the torsion disappears, the curvature remains. Nevertheless, by applying Weizenbock's connection, which is defined as \cite{Wanas1989,IWanas2007},

\begin{equation}\label{5a}
\Gamma^\Upsilon_{\mu\nu} = \lambda^\Upsilon_i \lambda_{i\mu},_{\nu}= -\lambda_{i\mu}, \lambda^{\Upsilon}_{i},_{\nu},
\end{equation}
where a non-symmetric link is represented by $\Gamma^\Upsilon_{\mu\nu}$. From this link, geometric objects can be obtained. The first is the torsion tensor, which has the following definition \cite{Wanas1989,IWanas2007}.

\begin{equation}
  \Lambda^\Upsilon{}_{\mu\nu}= \Gamma^\Upsilon{}_{\mu\nu} - \Gamma^\Upsilon{}_{\nu\mu}
= -\,\Lambda^\Upsilon{}_{\nu\mu}
\end{equation}
where $\Lambda^\Upsilon{}_{\mu\nu}$ is called the torsion tensor. It is now possible to define the third-order tensor as follows:
\begin{equation}
\Gamma^\Upsilon_{\mu\nu} = \{^{\Upsilon}_{\mu\nu}\} + \psi^\Upsilon_{\mu\nu}
\end{equation}
where $\psi^\Upsilon{}_{\mu\nu}$ denoted the contortion tensor.
\begin{equation}
\psi^\Upsilon{}_{\mu\nu}
= \lambda^\Upsilon_i \, \lambda_{i}{}_{\mu;\nu}
- \psi^\Upsilon_{\mu\nu},
\end{equation}
Eqs. \eqref{4a}–\eqref{5a} are utilized to establish the connection between the contortion tensor $\psi^\Upsilon{}_{\mu\nu}$ and the torsion tensor $\Lambda^\Upsilon{}_{\mu\nu}$ .
 
\begin{equation}\label{9a}
\psi_{\mu\nu\Upsilon} = \frac{1}{2}
\left( \Lambda_{\mu\nu\Upsilon} + \Lambda_{\nu\Upsilon\mu} + \Lambda_{\Upsilon\nu\mu} \right).
\end{equation}
Here is the torsion vector
\begin{equation}\label{10a}
\Lambda_{\mu} = \Lambda^{\nu}{}_{\mu\nu}
= - \Lambda^{\nu}{}_{\nu \mu}
= \psi^{\nu}{}_{\mu\nu},
\qquad
\psi^{\nu}{}_{\nu \mu} = 0.
\end{equation}

Physical applications commonly utilize these tensors. References \cite{MIWanas2009,Wanas2016} describe the general linear connection, as seen in eq. \eqref{10a}
\begin{equation}\label{11a}
\nabla^{\Upsilon}{}_{\mu\nu} = \{^{\Upsilon}{}_{\mu\nu}\} + \verb"A"\,\psi^{\Upsilon}{}_{\mu\nu}
\end{equation}
where \verb"A" is a dimensionless parameter. Using connection \eqref{11a} yields the Universal absolute derivative, which is \cite{Bakry2021}.

\begin{equation}
\mathscr{A}_{\mu\nu} = \mathscr{A}_{\mu, \nu} - \nabla^{\Upsilon}{}_{\mu\nu} \mathscr{A}_{\Upsilon}.
\end{equation}
The tensor of curvature is provided by 
\begin{equation}\label{13a}
B^{\epsilon}{}_{\mu\nu\Upsilon}
= \nabla^{\epsilon}{}_{\mu\Upsilon , \nu}
 - \nabla^{\epsilon}{}_{\mu\nu , \Upsilon}
 + \nabla^{k}{}_{\mu\Upsilon}\,\nabla^{\epsilon}{}_{k\nu}
 - \nabla^{k}{}_{\mu\nu}\,\nabla^{\epsilon}{}_{k\Upsilon}.
\end{equation}

Equation \eqref{11a} may be substituted into eq. \eqref{13a}, yielding

\begin{equation}\label{14a}
B^{\epsilon}{}_{\mu\nu\Upsilon}
= R^{\epsilon}{}_{\mu\nu\Upsilon}\{\}
+{\Sigma}^{\epsilon}{}_{\mu\nu\Upsilon},
\end{equation}
where $R^{\epsilon}{}_{\mu\nu\Upsilon}\{\}$ is the tensor produced by the Christoffel symbols, denoted by the following definition:

\begin{equation}
R^{\epsilon}{}_{\mu\nu\Upsilon}
= \{^{\epsilon}{}_{\mu\Upsilon}\}_{,\nu}
 - \{^{\epsilon}{}_{\mu\nu}\}_{,\Upsilon}
 + \{^{k}{}_{\mu\Upsilon}\}\{^{\epsilon}{}_{k\nu}\}
 - \{^{k}{}_{\mu\nu}\}\{^{\epsilon}{}_{k\Upsilon}\}.
\end{equation}

and 
\begin{equation}
{\Sigma}^{\epsilon}{}_{\mu\nu\Upsilon}
= \verb"A" \left( \psi^{\epsilon}{}_{\mu\Upsilon ; \nu}
- \psi^{\epsilon}{}_{\mu\nu ; \Upsilon} \right)
+ \verb"A"^{2} \left(
\psi^{k}{}_{\mu\Upsilon} \, \psi^{\epsilon}{}_{k\nu}
- \psi^{k}{}_{\mu\nu} \, \psi^{\epsilon}{}_{k\Upsilon}
\right).
\end{equation}
 The formulas for the generalised Ricci tensor are provided by eq. \eqref{14a}

\begin{equation}\label{17}
B_{\mu\Upsilon} = R_{\mu\Upsilon}\{\} + {\Sigma}_{\mu\Upsilon},
\end{equation}
where $R_{\mu\nu}$ is the Ricci tensor that is defined by 

\begin{equation}
R_{\mu\Upsilon}
=  \{^{\nu}{}_{\mu\Upsilon}\},_{\nu}
 -  \{^{\nu}{}_{\mu\nu}\},_{\Upsilon}
 + \{^{k}{}_{\mu\Upsilon}\}\{^{\nu}{}_{k\nu}\}
 - \{^{k}{}_{\mu\nu}\}\{^{\nu}{}_{k\Upsilon}\}
\end{equation}
and 
\begin{equation}\label{19}
{\Sigma}_{\mu\Upsilon}
= \verb"A"\left(\psi^{\nu}{}_{\mu\Upsilon;\nu}
          - \psi^{\nu}{}_{\mu\nu;\Upsilon}\right)
 + \verb"A"^{2}\left(
   \psi^{k}{}_{\mu\Upsilon}\,\psi^{\nu}{}_{k\nu}
 - \psi^{k}{}_{\mu\nu}\,\psi^{\nu}{}_{k\Upsilon}
 \right).
\end{equation}
The curvature scalar is therefore provided by 

\begin{equation}
B = g^{\mu\gamma} B_{\mu\Upsilon}= R\{\} + {\Sigma},
\end{equation}
 where
\begin{equation}
R= g^{\mu\Upsilon} R_{\mu\Upsilon}, \qquad {\Sigma} = g^{\mu\Upsilon} {\Sigma}_{\mu\Upsilon}.
\end{equation}

The term ``parameterized anticurvature tensor" \cite{Wanas2012} is used to describe Tensor \eqref{19}. It's worth noting that the Christoffel symbol, which appears in eq. \eqref{14a}, has established connections to gravitational forces. This suggests that gravity is the fundamental cause of space-time's curvature. Moreover, eq. \eqref{14a} demonstrates that ${\Sigma}^{\epsilon}_{\mu\nu\Upsilon}$  is a consequence of both torsion and contortion, as defined in eq. \eqref{9a}. For $\verb"A"=0$, the generalized Ricci tensor simplifies to the form given in eq. \eqref{17} by 
 
\begin{equation}
(B_{\mu\Upsilon})_{\verb"A"=0} = R_{\mu\Upsilon}{\{\}}.
\end{equation}
This case reduces to the gravitational field as in GR. For $\verb"A"=1$, the generalised Ricci tensor given in eq. \eqref{17} takes the form
\begin{equation}
(B_{\mu\Upsilon})_{\verb"A"=1} = R_{\mu\Upsilon}{\{\}} + (\Sigma_{\mu\Upsilon})_{\verb"A"=1} = 0.
\end{equation}
Thus,
\begin{equation}
(\Sigma_{\mu\Upsilon})_{\verb"A"=1} = -R_{\mu\Upsilon}{\{\}}.
\end{equation}
This case illustrates the flat space of gravity $\approx$ antigravity.
For $\verb"A"=-1$, the generalized Ricci tensor in eq. \eqref{17}
\begin{equation}
(B_{\mu\Upsilon})_{\verb"A"=-1} = 2 R_{\mu\Upsilon}{\{\}} + 2 \left( \psi^{k}{}_{\mu\Upsilon}\,\psi^{\nu}{}_{k\nu}
 - \psi^{k}{}_{\mu\nu}\,\psi^{\nu}{}_{k\Upsilon} \right).
\end{equation}
Torsion and a strong gravitational field are represented in this instance.
For $\verb"A"=2$, the generalized Ricci tensor takes the form given in eq. \eqref{17}, takes another specific form as given below.

\begin{equation}
(B_{\mu\Upsilon})_{\verb"A"=2} = - R_{\mu\Upsilon}{\{\}}+ 2 \left( \psi^{k}{}_{\mu\Upsilon}\,\psi^{\nu}{}_{k\nu}
 - \psi^{k}{}_{\mu\nu}\,\psi^{\nu}{}_{k\Upsilon} \right).
\end{equation}
 
The text describes an antigravitational field featuring torsion. It highlights that the parameter `\verb"A"' central to characterizing  the field's nature. When \verb"A" is zero, the field behaves as gravity in Riemann geometry. Conversely, for any non-zero value of \verb"A", the field operates under AP geometry, representing either gravity with torsion or antigravity with torsion, depending on that specific `\verb"A"' value. Fundamentally, `\verb"A"' acts as a modifier for the ratio of antigravity to gravity, defined as   
$\verb"A" = \text{antigravity / gravity}$.

\subsection{The $f (R, \Sigma, T)$ gravity and antigravity theory's mathematical formalism}

\hspace{0.5cm}The definition of the $f (R,\Sigma,T)$ theory's action principle is \cite{Bakry2023}.
It is convenient to start from the action
\begin{equation}\label{27}
I = \frac{1}{16\pi} \int_{\mathcal{D}} \sqrt{-g}\,\big[ f(R,\Sigma,T) + \mathcal{L}_m \big]\, d^4x,
\end{equation}
where \(\mathcal{D}\) is a four-dimensional spacetime region bounded by a closed three-surface, 
\(g = \det\mid g_{i\mu}\mid\), \(\mathcal{L}_m\) is the matter Lagrangian density, and 
\(f(R,\Sigma,T)\) is a function of the Ricci scalar \(R\), the (``antigravity'') Ricci scalar \(\Sigma\), 
and the trace of the energy--momentum tensor $T$. 
We adopt natural units where \(1 = G = c\).
Varying the eq. \eqref{27} with respect to \(g_{\mu\nu}\) gives the field equations as \cite{Bakry2023}:
\begin{equation}\label{28}
R_{\mu\nu}\, \frac{\partial f}{\partial R}
+ \Sigma_{\mu\nu}\, \frac{\partial f}{\partial \Sigma}
- \frac{1}{2} g_{\mu\nu} f
+ \left( g_{\mu\nu}\nabla^\Upsilon \nabla_\Upsilon - \nabla_\mu \nabla_\nu \right)
\left(
\frac{\partial f}{\partial R}
- \verb"A" \frac{\partial f}{\partial \Sigma} 
\right)
= 8\pi T_{\mu\nu} - \left( T_{\mu\nu} + \Theta_{\mu\nu} \right) \frac{\partial f}{\partial T},
\end{equation}
where \(\nabla_\mu\) denotes the covariant derivative compatible with \(g_{\mu\nu}\), and expansion $\Theta_{\mu\nu}$ defined as
\begin{equation}
\Theta_{\mu\nu} \equiv g_{\alpha\beta} \frac{\delta T^{\alpha\beta}}{\delta g_{\mu\nu}}.
\end{equation}

We assume that the matter content is a perfect fluid \cite{Brans1961},
\begin{equation}
T_{\mu\nu} = ( \rho + P ) u_\mu u_\nu - P g_{\mu\nu}, 
\qquad
T = \rho - 3P,
\qquad
\mathcal{L}_m = -P,
\qquad
u^\mu u_\mu = 1,
\qquad
u^\mu u_{\mu;\nu} = 0,
\end{equation}
where \(\rho\) is the energy density, \(P\) the pressure, and \(u^\mu\) the four-velocity of the fluid in
co-moving coordinates. Under these assumptions, the field equation \eqref{28} reduce to \cite{Bakry2023}
\begin{equation}
R_{\mu\nu}\, \frac{\partial f}{\partial R}
+ \Sigma_{\mu\nu}\, \frac{\partial f}{\partial \Sigma}
- \frac{1}{2} g_{\mu\nu} f
+ \left( g_{\mu\nu}\nabla^\Upsilon \nabla_\Upsilon - \nabla_\mu \nabla_\nu \right)
\left(
\frac{\partial f}{\partial R}
- \verb"A" \frac{\partial f}{\partial \Sigma}
\right)
= 8\pi T_{\mu\nu} + \left( T_{\mu\nu} + P g_{\mu\nu} \right) \frac{\partial f}{\partial T}
\end{equation}

In this work we choose the specific functional form
\begin{equation}\label{32}
f(R,\Sigma,T) = R + \Sigma + 2 \aleph T,
\end{equation}
where $\aleph$ is a constant parameter.

Using the function \eqref{32}, the final form of the field equations of $f(R,\Sigma,T)$ is
\begin{equation}\label{33}
R_{\mu\nu} + \Sigma_{\mu\nu} - \frac{1}{2} g_{\mu\nu} (R + \Sigma)
= 2(4\pi + \aleph)\, T_{\mu\nu} + \aleph\, g_{\mu\nu}(T + 2P),
\end{equation}
where
\begin{equation}\label{34}
G^{*}_{\mu\nu} \equiv R_{\mu\nu} + \Sigma_{\mu\nu} - \frac{1}{2} g_{\mu\nu} (R + \Sigma).
\end{equation}

Several key special cases are worth examining:
\begin{itemize}
    \item When $\verb"A" = 0$, the $f(R,\Sigma,T)$ field equation \eqref{33} simplify to the  $f(R,T) = R + 2\aleph T$ gravity model, as explored by \cite{Harko2011}.
    \item For $\verb"A" = \aleph = 0$, the eq.~\eqref{33} reduce to $f(R)$ gravity, specifically with $f(R) = R$, corresponding to standard GR, as discussed by \cite{Santos2007}.
    \item For $\verb"A" = 1$, the eq.~\eqref{33} yield a scenario akin to antigravity or a flat space within the $f(R,\Sigma,T)$ gravity framework.
    \item When $\verb"A" < 0$, the $f(R,\Sigma,T)$ gravity model corresponds to a strong gravitational field~\eqref{33}.
    \item For $\verb"A" > 1$, the field equations~\eqref{33} transition into an antigravity regime within the $f(R,\Sigma,T)$ gravity framework.
\end{itemize}

The non-conservation of the energy–momentum tensor is
\begin{equation}\label{35c}
\nabla^\mu T_{\mu\nu}
= - \frac{\aleph}{2(4\pi + \aleph)} \,
\nabla^\mu \!\left[ g_{\mu\nu} (T + 2P) \right].
\end{equation}
One may verify that the energy–momentum tensor is maintained, as in the case of GR theory, by changing $\aleph = 0$ in eq.~\eqref{35c}, Additionally, the energy–momentum tensor is not preserved if $\verb"A" = 0$, as is the case with $f(R,T)$ gravity.
The upcoming section will focus on deriving key quantities to facilitate solving the $f(R,\Sigma,T)$  field equations, likely involving mathematical expressions and calculations to gain insight into the model's behavior.

In the next section, we will calculate quantities that help us to solve the $f(R,\Sigma,T)$ field equations.

\section{Tetrad fields as orthonormal frames}

\hspace{0.5cm}A spherically symmetric geometry in curvature coordinates can be described using a specific tetrad field adapted to those coordinates.
\((t,r,\theta,\phi)\),
\begin{equation}\label{36}
\lambda^{\mu}{}_{i} = \mathrm{diag}\!\left( e^{-\nu},\, e^{-\lambda},\, \frac{1}{r},\, \frac{1}{r\sin\theta} \right),
\end{equation}
where \(\nu=\nu(r)\) and \(\lambda=\lambda(r)\) are undetermined functions of \(r\). Using \eqref{1a} and \eqref{36}, one gets the inverse tetrad
\begin{equation}\label{37}
\lambda_{i \mu} = \mathrm{diag}\!\left( e^{\nu},\, e^{\lambda},\, r,\, r\sin\theta \right).
\end{equation}

Substituting \eqref{37} into \eqref{3a}, one obtains the metric tensor
\begin{equation}\label{38}
g_{\mu\nu} = \mathrm{diag}\!\left( e^{2\nu},\, -e^{2\lambda},\, -r^{2},\, -r^{2}\sin^{2}\theta \right).
\end{equation}

Hence, using \eqref{38} into \eqref{2a} we get the line element
\begin{equation}\label{39}
ds^{2} = e^{2\nu} dt^{2} - e^{2\lambda} dr^{2} - r^{2}\left(d\theta^{2} + \sin^{2}\theta\, d\phi^{2}\right).
\end{equation}

Note that if \(\nu\) and \(\lambda\) vanish (e.g. at some point or everywhere), the metric \eqref{39} reduces to the Minkowski metric in spherical coordinates. The non-zero Ricci tensor components are
\begin{equation}
\begin{aligned}
B_{00} &= (1-\verb"A")\, e^{2(\nu-\lambda)}
\left( \nu'' + (\nu')^{2} - \lambda' \nu' + \frac{2(1-\texttt{A})\,\nu'}{r} \right), \\
B_{11} &= (\verb"A"-1)\left( \nu'' + (\nu')^{2} - \nu' \lambda' - \frac{2\lambda'}{r} \right),
\end{aligned}
\end{equation}
\begin{equation}
\begin{aligned}
B_{22} &= (1-\verb"A")\, e^{-2\lambda} \left[ (\verb"A"-1)\, r \nu' + r \lambda' + (\verb"A"-1) \right] + (1-\verb"A"), \\
B_{33} &= \sin^{2}\theta \, B_{22}.
\end{aligned}
\end{equation}

And,
\begin{equation}
\begin{aligned}\label{42}
G^{*}{}_{0}{}^{0} &= (1-\verb"A")\, \frac{e^{-2\lambda}}{r^{2}} \left( e^{2\lambda} + 2 r \lambda' + \verb"A" - 1 \right), \\
G^{*}{}_{1}{}^{1} &= (1-\verb"A")\, \frac{e^{-2\lambda}}{r^{2}}
\left( e^{2\lambda} + 2 (\verb"A"-1) r \nu' + \verb"A" - 1 \right),
\end{aligned}
\end{equation}
\begin{equation}
\begin{aligned}
G^{*}{}_{2}{}^{2} &= (1-\verb"A")\, e^{-2\lambda} \left[
\frac{\lambda'}{r} + (\verb"A"-1)\frac{\nu'}{r} - \left( \nu'' + (\nu')^{2} - \lambda' \nu' \right)
\right], \\
G^{*}{}_{3}{}^{3} &= \sin^{2}\theta \, G^{*}{}_{2}{}^{2}.
\end{aligned}
\end{equation}

The energy–momentum tensor is
\begin{equation}\label{44}
T^{0}{}_{0} = \rho, \qquad
T^{1}{}_{1} = T^{2}{}_{2} = T^{3}{}_{3} = -P.
\end{equation}

We will examine the three regions of a gravastar in the next section.

\section{Gravastars will be studied in the $f(R,\Sigma,T)$ gravity framework using the field equations.}
\hspace{0.5cm}In this section, we implement and analyze the gravaster model: The gravaster model consists of an interior region  \(0 \le r < r_1\) surrounded by a thin shell of ultrarelativistic  matter  \(r_1 \le r < r_2\), where the shell's thickness 
 \(r_2 = r_1 + \varepsilon\) is extremely small, the exterior region   \(r_2 \le r < r_3\) is a vacuum, making the Schwarzschild spacetime a suitable description for this part of the gravaster model.
The system is divided into three parts based on the EoS. The interior region has an EoS \(P = -\rho\), characteristic of dark energy, with \(\verb"A" = 2\), indicating a repulsive, antigravity nature due to torsion. The shell has an EoS \(P = \rho\), indicating ultrarelativistic matter, with \(\verb"A" = -1\), suggesting strong gravity with torsion.
The exterior region has $P = \rho = 0$, indicating a vacuum, with  $(\verb"A" = 0)$, meaning gravity without torsion, consistent with the Schwarzschild spacetime description.
Substituting eqs.~\eqref{42}–\eqref{44} into eq. \eqref{33}, we obtains
\begin{equation}\label{45}
(1-\verb"A")\,\frac{e^{-2\lambda}}{r^{2}}
\left( e^{2\lambda} + 2 r \lambda' + \verb"A" - 1 \right)
= 8\pi \rho + \aleph (3\rho - P),
\end{equation}
\begin{equation}
(1-\verb"A")\,\frac{e^{-2\lambda}}{r^{2}}
\left( e^{2\lambda} + 2 (\verb"A"-1) r \nu' + \verb"A" - 1 \right)
= - 8\pi P + \aleph (\rho - 3P),
\end{equation}
and
\begin{equation}\label{47}
(1-\verb"A")\, e^{-2\lambda} \left[
\frac{\lambda'}{r} + (\verb"A"-1)\frac{\nu'}{r}
- \left( \nu'' + (\nu')^{2} - \lambda' \nu' \right)
\right]
= - 8\pi P + \aleph (\rho - 3P).
\end{equation}
Equation~\eqref{35c} implies the non-conservation of the energy-momentum tensor.
\begin{equation}\label{48}
\frac{dP}{dr} + \nu' (P + \rho)
+ \frac{\aleph}{2(4\pi + \aleph)} \left( \frac{dP}{dr} - \frac{d\rho}{dr} \right) = 0.
\end{equation}
When \(\verb"A"= 0\), eqs.~\eqref{45}–\eqref{48} simplify to the standard gravastar equations in general relativity or a specific modified gravity context, likely $f(R,T)$ gravity.

As a special case for the metric potential $e^{2v}$, we have assumed the Kuchowicz-type, 

\begin{equation}
e^{2v}= e^{Br^2+2log[C]}
\end{equation}
The metric potential, involving constants $B$ and $C$, is well behaved and singularity-free, with B having dimension $[L^{-2}]$ and $C$ being dimensionless.\\
Substituting the condition into eqs. \eqref{45}–\eqref{47} yield the following equation:
\begin{equation}\label{50}
(1-\verb"A")\,\frac{e^{-2\lambda}}{r^{2}}
\left( e^{2\lambda} + 2 r \lambda' + \verb"A" - 1 \right)
= 8\pi \rho + \aleph (3\rho - P),
\end{equation} 
\begin{equation}\label{51}
(1-\verb"A")\,\frac{e^{-2\lambda}}{r^{2}}
\left( e^{2\lambda} + 2 (\verb"A"-1)Br^{2}  + \verb"A" - 1 \right)
= - 8\pi P + \aleph (\rho - 3P),
\end{equation}
 
 \begin{equation}\label{52}
(1-A)\, e^{-2\lambda} \frac{1}{r} 
\left[ \lambda' + b B r - 2 B r - B^{2} r^{3} + \lambda' B r^{2} \right]
= - 8 \pi P + \aleph (\rho - 3P).
\end{equation}

\section{Interior spacetime}

\hspace{0.5cm}Following Mazur–Mottola’s proposal, we assume that the EoS inside the region is
\begin{equation}\label{53}
p = - \rho.
\end{equation}
This is a special case of \( p = \omega \rho \) with \( \omega = -1 \) and $\verb"A"=2$, which is known as the dark energy EoS. From eq.~\eqref{53}, we obtain
\begin{equation}
\rho = \rho_0 \; (\text{constant}),
\end{equation}
and the pressure becomes
\begin{equation}\label{55}
p = - \rho_0.
\end{equation}
Using the eq. \eqref{55} in eqns. \eqref{50}-\eqref{52}, we obtain

\begin{equation}\label{56}
\frac{-e^{-2\lambda}}{r^{2}}
\left( e^{2\lambda} + 2 r \lambda' + 1 \right)
= 8\pi \rho_{0} + 4\aleph\rho_{0},
\end{equation} 
\begin{equation}
\frac{-e^{2\lambda}}{r^{2}}
\left( e^{2\lambda} + 2Br^{2}  + 1 \right)
= 8\pi \rho_{0} + 4\aleph\rho_{0},
\end{equation}
 
 \begin{equation}
 -e^{-2\lambda} \frac{1}{r} 
\left[ \lambda' + 2B r - 2 B r - B^{2} r^{3} + \lambda' B r^{2} \right]
= 8\pi \rho_{0} + 4\aleph\rho_{0}.
\end{equation}
Equation \eqref{56} is rewriten as 
\begin{equation}\label{59}
2 r \lambda' \, e^{-2\lambda} + e^{-2\lambda} + 1 = -4 (2\pi + \aleph)\, \rho_0 \, r^2 .
\end{equation}

Integrating \eqref{59}, we get
\begin{equation}
e^{-2\lambda} = 4 (2\pi + \aleph)\, \rho_0 \, r^2 + A r - 1,
\end{equation}
where \(A\) is an integration constant. We set \(A = 0\) to maintain the solution regular at the centre (\(r = 0\)). Consequently,
\begin{equation}
e^{-2\lambda} = 4 (2\pi + \aleph)\, \rho_0 \, r^2 - 1,
\end{equation}

\begin{figure}[h!]
  \centering
  \begin{minipage}[b]{0.6\textwidth}
    \centering
   \includegraphics[width=0.9\textwidth]{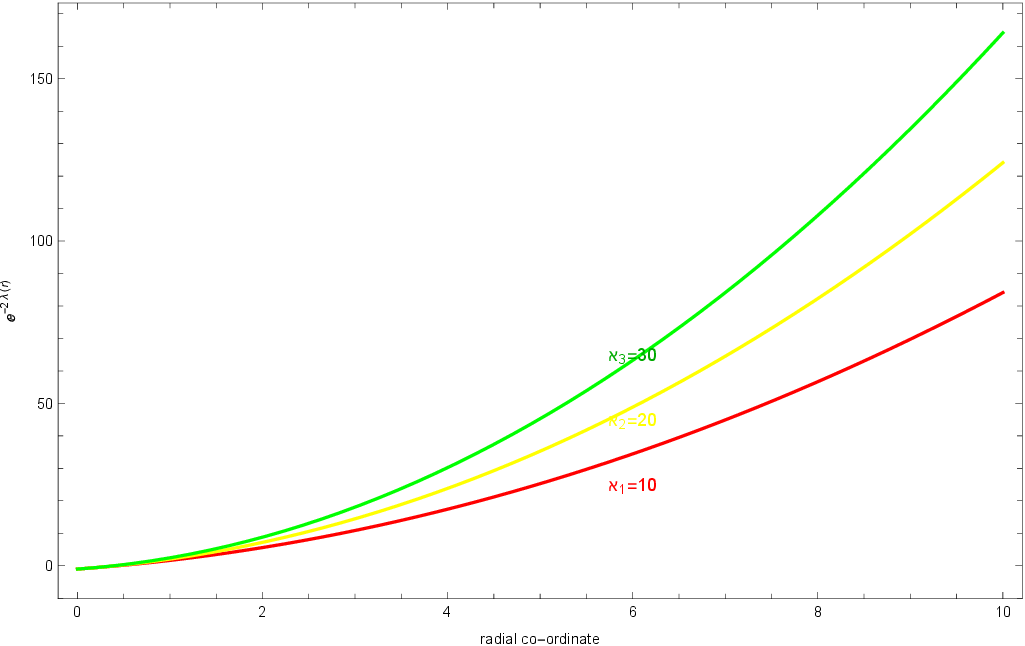}
  \caption{The interior region's metric potential is plotted as a function of radial distance r(km) for $\rho_{0}=0.01$.}
   \label{1}
  \end{minipage}
  \hfill
  \begin{minipage}[b]{0.8\textwidth}
    \centering
   \includegraphics[width=0.57\textwidth]{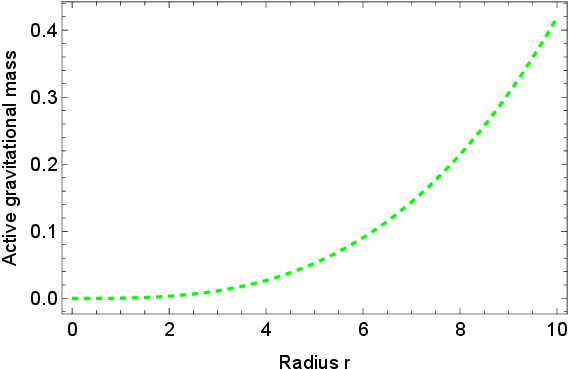}
  \caption{ $\texttt{M(d)}$ is plotted against radial distance r(km).}
   \label{2}
  \end{minipage}
\end{figure}
The above results clearly show that there is no singularity in the inner solutions, thereby solving the core singularity problem of a classical black hole. To provide further clarity, we have displayed in Fig. \ref{1} the fluctuation of the metric potential $e^{-2\lambda}$
with respect to the radial parameter $r$. Additionally, the gravitational mass $\texttt{M(d)}$ is determined by
\begin{equation}
\texttt{M(d)} = \int_{0}^{r_1 = d} 4\pi r^{2}\rho_0 \, dr
      = \frac{4}{3}\pi d^{3}\rho_0.
\end{equation}
\section{Shell}
\hspace{0.5cm}We consider a stiff perfect fluid inside the thin shell with the equation of state 
\begin{equation}\label{63}
  p=\rho
\end{equation}
which is a special case of the barotropic form $p=\omega\rho$, for $\omega=1$ and $\verb"A"=-2$. In barotropic fluids, pressure depends only on density. Though uncommon in nature, their simplicity helps in exploring various physical scenarios. Zel’dovich first introduced stiff fluids in connection with a cold baryonic universe, and studied their role in spherical collapse  \cite{Staelens2021}. This model has since been widely applied in astrophysics and cosmology \cite{Rahaman2014,Meghanil2025,Sinha}. The field equations in the shell are difficult to solve, but under the thin-shell approximation  
$0<e^{-2(\lambda)}\leq1$  an analytical solution is possible. As per Israel \cite{Israel1966}, the region between two spacetimes must form a thin shell, and parameters depending on $r$
can be treated as negligible as $r\rightarrow0$. With this, the field eqs. \eqref{50}–\eqref{52} simplify to the following equations.

\begin{equation}\label{64}
 \frac{1}{r^2}+\frac{2r\lambda' e^{-2\lambda}}{r^2}-\frac{3e^{-2\lambda}}{r^2}=\frac{1}{3}(8\pi+2\aleph)\rho
\end{equation}
\begin{equation}\label{65}
 \frac{1}{r^2}-6B{e^{-2\lambda}}-\frac{3e^{-2\lambda}}{r^2}=-\frac{1}{3}(8\pi+2\aleph)\rho
\end{equation}
\begin{equation}\label{66}
 \frac{e^{-2\lambda}}{r}(\lambda'-4Br-B^{2}r^{3}+\lambda'Br^{2})=-\frac{1}{3}(8\pi+2\aleph)\rho
\end{equation}
and 
\begin{equation}\label{67}
  \frac{d\rho}{dr} + B r( 2\rho) = 0
\end{equation}

By utilizing eq. \eqref{63} in eqs. \eqref{64} and \eqref{65}, and using thin shell condition, we obtain the metric potential as

\begin{figure}[h!]
  \centering
  \includegraphics[width=0.37\textwidth]{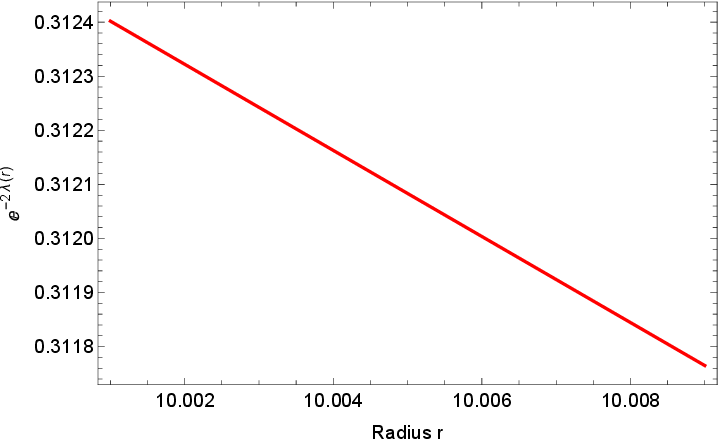}
  \caption{The metric potential of the shell is shown to with its thickness, measured in km for $H_{1}=0.002,B=0.01491932683$.}
   \label{3}
\end{figure}

\begin{equation}
  e^{-2\lambda}=e^{-3 B r^{2}} H_{1} + e^{-3 B r^{2}} \, \mathrm{ExpIntegralEi}(3B r^{2})
\end{equation}
where $H_{1}$ is integration constant. The change in the metric potential is seen Fig. \ref{3}. It is clear it grows monotonically as one gets closer to the outer limit of the shell. We can thus physically demonstrate that our solution is regular and singularity-free. As per eq. \eqref{67}, we obtain the pressure and matter density for the intermediate shell region which is given as
\begin{equation}\label{69F}
  \rho = P =  e^{-Br^2}K_{1}
\end{equation}
Here, $K_{1}$ represents a constant.
\section{Exterior spacetime}
\hspace{0.5cm}The exterior region, where pressure and density are zero $(p = \rho = 0)$, is described by the static exterior Schwarzschild solution.
\begin{equation}
  ds^{2} = \left( 1 - \frac{2M}{r} \right) dt^{2}
 - \left( 1 - \frac{2M}{r} \right)^{-1} dr^{2}
 - r^{2} \left( d\theta^{2} + \sin^{2} \theta \, d\phi^{2} \right).
 \label{68}
\end{equation}
Here, $M$ represents the total mass of the system generating the gravitational field.
 
\section{Certain physical properties of the shell}

\subsection{Proper length of the shell}
\hspace{0.5cm}The shell exists where two spacetimes meet, as proposed by Mazur and Mottola \cite{Mazur2023,Mazur2004}. The shell's length spans between two boundaries: the outer edge  $(r_{2} = d + \varepsilon)$ and the inner edge $r_{1}=d$. The shell's thickness is determined by the region between these boundaries. \\
\[
\ell = \int_{d}^{d+\epsilon} \sqrt{{e^{2\lambda}}}\, dr
\]
\begin{equation}\label{69}
  = \int_{d}^{d+\epsilon} \frac{dr}{\sqrt{e^{-3 B r^{2}} H_{1} + e^{-3 B r^{2}} \, \mathrm{ExpIntegralEi}(3 B r^{2})}}
\end{equation}
\[
=\int_{d}^{d+\epsilon} \frac{dr}{f(r)},
\]
with
\begin{equation}
f(r) = \sqrt{e^{-3 B r^{2}} H_{1} + e^{-3 B r^{2}} \, \mathrm{ExpIntegralEi}(3 B r^{2})}.
\end{equation}
To simplify the integral in eq. \eqref{69}, we choose $\frac{df(r)}{dr} = \frac{1}{f(r)}$  which makes it easier to solve. Thus, we obtain\\
 \begin{equation}\label{71}
   l = f(d + \epsilon) - f(d)
 \end{equation}
Expanding in a Taylor series and keeping only linear terms in $\varepsilon$, we get from eq. \eqref{71} 
\begin{equation}
  l = \epsilon \,\frac{df(r)}{dr}
   \approx \epsilon \sqrt{e^{-3 B r^{2}} H_{1} + e^{-3 B r^{2}} \, \mathrm{ExpIntegralEi}(3 B r^{2})
   }.
\end{equation}
\begin{figure}[h!]
  \centering
  \includegraphics[width=0.4\textwidth]{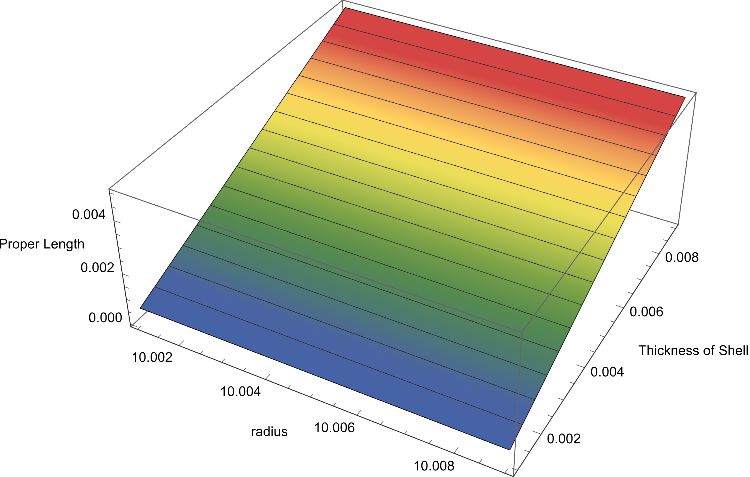}
  \caption{The proper length of the shell varies with its thickness, measured in km for $H_{1}=0.002,B=0.01491932683$.}
   \label{4}
\end{figure}
Given $\varepsilon$ small value, higher order terms in the exponential are negligible. Figure \ref{4} illustrates the proper length variation for the thin shell's radius.
The relation between proper length relevant parameters is depicted in figure \ref{4}, a gradual increase is observed in the plot of proper length versus shell thickness and radius. This suggests $f(R,\Sigma,T)$ gravity contributes to the shell's length increase in the variable case.

\subsection{Entropy}
\hspace{0.5cm}Following Mazur and Motola's work \cite{Mazur2004}, the interior region has zero entropy density, indicating a single condensate state. Within the shell, however, the entropy is defined differently as
\begin{equation}\label{73}
S = \int_{d}^{d+\epsilon} 4\pi r^{2}\, s(r)\, \sqrt{e^{\lambda}} \, dr .
\end{equation}
The relationship between entropy density \( s(r) \) and local temperature \( T(r) \) is given by a known formula ~\cite{P.Mazur,Mottola2004}.
\begin{equation} \label{eq:45}
s(r) = \frac{\alpha^2 k_B^2}{4\pi \hbar^2 T(r)} 
     = \alpha \left( \frac{k_B}{\hbar} \right) \sqrt{\frac{P}{2\pi}},
\end{equation}
Here, \( \alpha \) represents a constant.
We use geometrized units \( G = c = 1 \) and Planckian units \( k_B = \hbar = 1 \), simplifying the entropy density within the shell.
\begin{equation} \label{eq:46}
s(r) = \alpha \sqrt{\frac{P}{2\pi}}.
\end{equation}
Eq. \eqref{73} then takes the from
\begin{equation}\label{76}
S = \int_{d}^{d+\epsilon} 4\pi r^2 \, \alpha
\left( \frac{P}{2\pi} \right)^{\frac{1}{2}} \sqrt{e^{2\lambda}} \, dr.
\end{equation}
where
\begin{equation}\label{77}
N = \int_{d}^{d+\epsilon} D(r) \, dr,
\end{equation}
\begin{equation}
D(r)=\sqrt{\frac{e^{-Br^{2}}{r^{4}} }{e^{-3 B r^{2}} H_{1} + e^{-3 B r^{2}} \, \mathrm{ExpIntegralEi}(3 B r^{2})}}
\end{equation}
To tackle eq. \eqref{77}, we find the primitive F(r) of D(r) and apply the fundamental theorem of calculus to simplify it.
\begin{equation}\label{79}
N = \left[ F(r) \right]_{d}^{d+\epsilon} = F(d + \epsilon) - F(d).
\end{equation}
Expanding in a Taylor series around ``d" and keeping only liner term in $\varepsilon$, we derive a result from eq. \eqref{76} based on eq. \eqref{79}.
\begin{equation}
S = 2 \sqrt{2\pi} \, \alpha \sqrt{K_{1}} \, \varepsilon \, 
\sqrt{ \frac{e^{-B r^{2}} r^{4}}{e^{-3 B r^{2}} H_{1} + e^{-3 B r^{2}} \, \mathrm{ExpIntegralEi}(3 B r^{2})} }
\end{equation}
\begin{figure}[h!]
  \centering
  \includegraphics[width=0.4\textwidth]{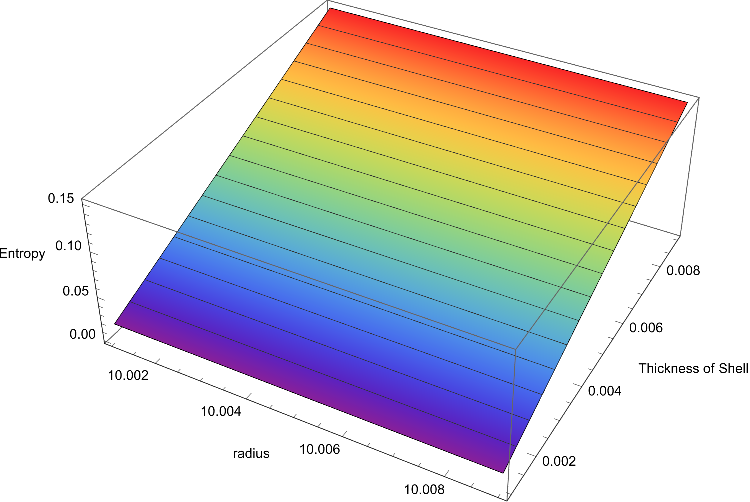}
  \caption{The entropy of the shell changes with its thickness, measured in km for $\alpha=0.2,K_{1}=0.5,H_{1}=0.002,B=0.01491932683$.}
   \label{5}
\end{figure}
We obtained the entropy expression for our model, with its variation shown in Figure \ref{5}. As shown in Figure \ref{5}, entropy increases progressively with the radial coordinate and thickness. The entropy reaches its maximum value on the gravastar's surface, ensuring physical consistency.

\subsection{Energy content}
\begin{figure}[h!]
  \centering
  \includegraphics[width=0.5\textwidth]{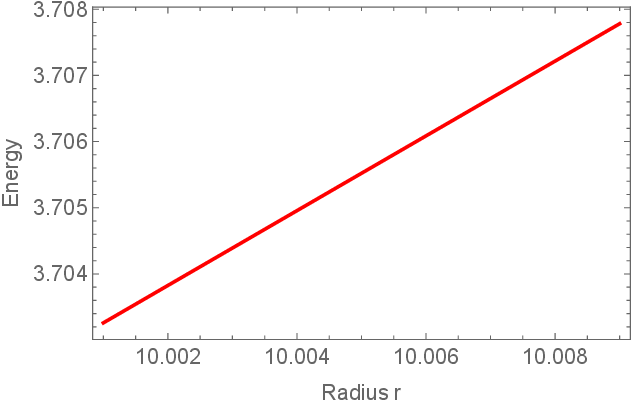}
  \caption{The energy of the shell varies with its thickness, measured in km for $K_{1}=0.002,B=0.01491932683$.}
   \label{6}
\end{figure}

\hspace{0.5cm}The shell's energy is calculated using a specific formula.
\begin{equation}\label{83}
E = \int_{d}^{d+\epsilon} 4\pi \rho \, r^2 \, dr
\end{equation}
 Combining eqs. \eqref{69F} and \eqref{83} gives
\begin{equation}\label{82}
E = \int_{d}^{d+\epsilon} 4\pi r^2 e^{-Br^{2}} K_{1} \, dr
\end{equation} 
Integrating eq. \eqref{82}, we get
\begin{equation}
E= 4 K_{1} \pi \left( -\frac{e^{-B r^2} r}{2B} + \frac{\sqrt{\pi} \, \mathrm{Erf}(\sqrt{B} \, r)}{4 B^{3/2}}\right)_{d}^{d+\epsilon}
\end{equation} 
There's clear link between energy and shell thickness, as illustrated in Figure \eqref{6}.
The evolution of the shell energy is seen in Fig. \ref{6}. This graph demonstrates that energy increases with shell thickness. It meets the requirement that the energy of the shell must increase as the radial distance increases.

\section{Junction condition}
\hspace{0.5cm}A gravaster has three regions interior (I), shell (II) and exterior (III). The interior and exterior regions meet at the shell's junction interface, requiring smooth matching between regions I and III, as per the Darmois-Israel formalism \cite{Darmois1927}.
At the junction surface $\Sigma$,i.e., at $r=D$ the metric coefficients are continuous, but their derivatives might not be. Using the Darmois-Israel formalism, we can determine the surface stress-energy tensor $S_{ij}$.
The Lanczos equations \cite{Lanczos1924} define the intrinsic surface stress-energy tensor  $S_{ij}$.
\begin{equation}
S^{i}_{\; j} = -\frac{1}{8\pi} \left( \kappa^{i}_{\; j} - \delta^{i}_{\; j} \, \kappa^{k}_{\; k} \right)
\end{equation}
Here, $\kappa_{ij} = K^{+}_{ij} - K^{-}_{ij}$ measures the difference in extrinsic curvature of the shell. This difference represents a discontinuity in the second fundamental forms \cite{Rahaman2006,Rahaman2011}. 

\begin{equation}
K_{ij} = -n_{\nu} \left( 
\frac{\partial^{2} x^{\nu}}{\partial \xi^{i} \partial \xi^{j}} 
+ \Gamma^{\nu}_{\alpha\beta} 
\frac{\partial x^{\alpha}}{\partial \xi^{i}} 
\frac{\partial x^{\beta}}{\partial \xi^{j}} 
\right) \Bigg|_{\Sigma}
\end{equation}

\noindent
Here, \( \xi^{i} \) are the intrinsic coordinates on the shell (i.e., coordinates defined on the hypersurface \( \Sigma \), and \( n_{\nu} \) are the components of the unit normal vector to the surface \( \Sigma \)). Spherically symmetric and static spacetime can be written as
\begin{equation}
ds^2 = f(r)\, dt^2 - \frac{1}{f(r)}\, dr^2 - r^2 \left( d\theta^2 + \sin^2\theta\, d\phi^2 \right),
\end{equation}
where \( f(r) \) is a generic function of the radial coordinate \( r \). The unit normal vector \( n_{\nu} \) to the surface \( \Sigma \) can be written as
\begin{equation}
n_{\nu} = 
\left( g^{\alpha\beta} \frac{\partial f}{\partial x^{\alpha}} \frac{\partial f}{\partial x^{\beta}} \right)^{-\frac{1}{2}} 
\frac{\partial f}{\partial x^{\nu}}, \label{eq:33}
\end{equation}
with the normalization condition \( n^{\mu} n_{\mu} = 1 \). Using the Lanczos equation, the surface stress-energy tensor on the shell can be written as
\[
S^{i}_{\; j} = \text{diag}[\sigma,\ -\upsilon,\ -\upsilon,\ -\upsilon],
\]
where \( \sigma \) is the surface energy density and \( \upsilon \) is the surface pressure. These are given by:

\begin{equation}
\sigma = -\frac{1}{4\pi d} \left[ \sqrt{f} \right]^{+}_{-}, \label{eq:34}
\end{equation}

\begin{equation}
\upsilon = -\frac{\sigma}{2} + \frac{1}{16\pi} \left[ \frac{f'}{\sqrt{f}} \right]^{+}_{-}, 
\end{equation}

Substituting the explicit forms of \( f(r) \), we obtain:

\begin{equation}\label{90}
\sigma = -\frac{1}{4\pi d} \left[ 
\sqrt{1 - \frac{2M}{d}} - 
\sqrt{{4(2\pi + \aleph)\rho_0 d^2+Ad-1}} 
\right],
\end{equation}

\begin{equation}\label{91}
\upsilon = \frac{1}{8\pi d} \left[
\sqrt{1 - \frac{2M}{d}} 
- \sqrt{-1 + A d + 4\rho_{0}(\aleph + 2\pi) d^2} 
+ \frac{d}{2} \left(
\frac{2M}{d^2 \sqrt{1 - \frac{2M}{d}}} 
- \frac{A + 8\rho_{0}(\aleph + 2\pi) d}{\sqrt{-1 + A d + 4\rho_{0}(\aleph + 2\pi) d^2}}
\right)
\right]
\end{equation}

\begin{figure}[h!]
  \centering
  \begin{minipage}[b]{0.6\textwidth}
    \centering
   \includegraphics[width=0.75\textwidth]{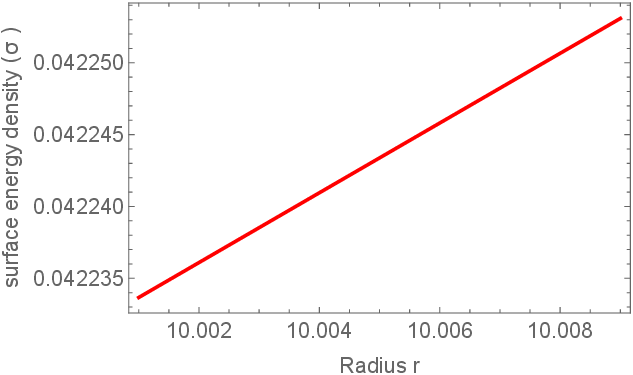}
  \caption{Variation of the surface energy density with respect to the radial co-ordinate(km) for $M = 0.338m_{\bigodot}, A=-2.5,\rho_{0}=0.01,\aleph=10$.}
\label{7}
  \end{minipage}
  \hfill
  \begin{minipage}[b]{0.39\textwidth}
    \centering
    \includegraphics[width=1.05\textwidth]{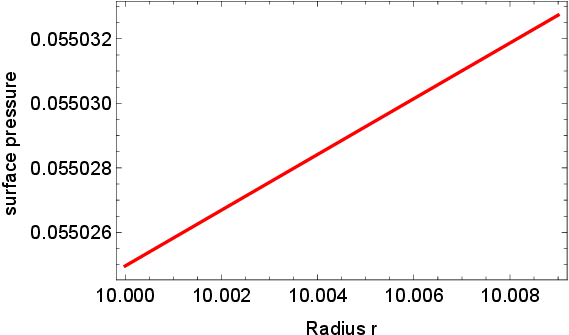}
  \caption{Variation of the surface pressure with respect to the radial co-ordinate(km) for $M = 0.338m_{\bigodot}, A=-2.5,\rho_{0}=0.01,\aleph=10$.}
\label{8}
  \end{minipage}
\end{figure}

The mass of the thin shell can then be written as
\begin{equation}
m_s = 4\pi d^2 \sigma = -d\left[ 
\sqrt{1 - \frac{2M}{d}} - 
\sqrt{{4(2\pi + \aleph)\rho_0 d^2+Ad-1}} 
\right],
\end{equation}

Here, \( M \) is the total mass of the gravastar.

\begin{equation}
M=\frac{-m_s{^2} + 2 d^2 - A d^3 - 4 \aleph \rho_{0} d^4 - 8 \rho_{0} \pi d^4 + 2m_s{^2} d\sqrt{-1 + A d + 4 \aleph \rho_{0} d^2 + 8 \rho_{0} \pi d^2}}{2 d}
\end{equation}
Figures \ref{7} and \ref{8} display the surface energy density and surface pressure as a function of $r$. It is evident that both values remain positive throughout the shell, proving that the null energy condition required to construct a thin shell model has been satisfied.

\section{Equation of state parameter}
\hspace{0.5cm}The EoS parameter at $r=d$ can be written as \cite{Pradhan2023}:
\begin{equation}\label{96}
 \omega(d)=\frac{\upsilon}{\sigma}
\end{equation}
Using eqs. \eqref{90} and \eqref{91} in \eqref{96}, the EoS parameter can be explicitly expressed. 
\begin{equation}
   \omega(d)=\frac{\frac{1}{8\pi d} \left[
\sqrt{1 - \frac{2M}{d}} 
- \sqrt{-1 + A d + 4\rho_{0}(\aleph + 2\pi) d^2} 
+ \frac{d}{2} \left(
\frac{2M}{d^2 \sqrt{1 - \frac{2M}{d}}} 
- \frac{A + 8\rho_{0}(\aleph + 2\pi) d}{\sqrt{-1 + A d + 4\rho_{0}(\aleph + 2\pi) d^2}}
\right)
\right]}{ -\frac{1}{4\pi d} \left[ 
\sqrt{1 - \frac{2M}{d}} - 
\sqrt{{4(2\pi + \aleph)\rho_0 d^2+Ad-1}} 
\right]}
\end{equation}

\begin{figure}[h!]
  \centering
  \includegraphics[width=0.4\textwidth]{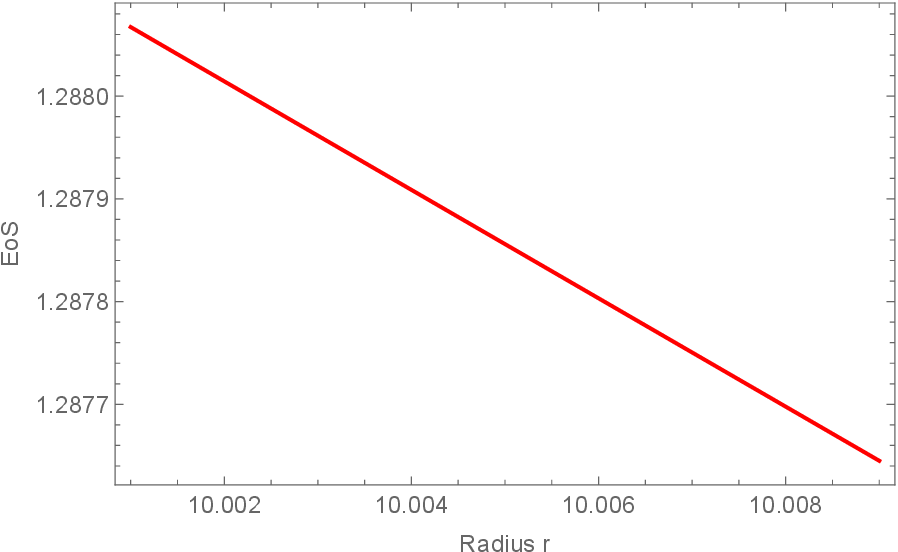}
  \caption{The EoS varies with respect to r for $M = 0.338m_{\bigodot}, A=-2.5,\rho_{0}=0.01,\aleph=10$.}
   \label{9}
\end{figure}
From the figure \ref{9} we can clearly see that the EoS parameter stays positive and decreases with radial co-ordinate towards the outer region featuring a feasible gravastar model.

\section{Stability}
\subsection{Speed of sound analysis}
\hspace{0.5cm}We've calculated pressure $(\upsilon)$ and energy density $(\sigma)$ for the thin shell using eqs. \eqref{90} and \eqref{91}. The speed of sound in this fluid can be expressed as:

\begin{equation}
\eta = \frac{\upsilon'}{\sigma'}
\end{equation}

\begin{figure}[h!]
  \centering
  \includegraphics[width=0.4\textwidth]{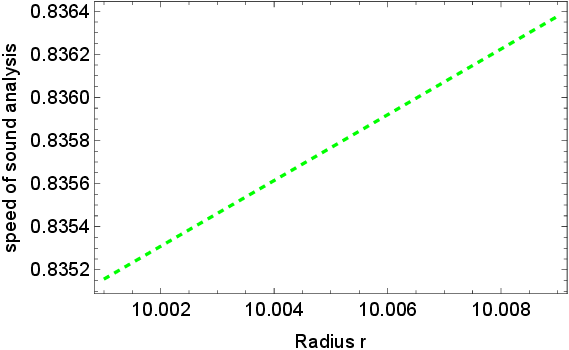}
  \caption{The stability of the shell respect to the radial co-ordinate(km) $M = 0.338m_{\bigodot}, A=-2.5,\rho_{0}=0.01,\aleph=10$.}
  \label{10}
\end{figure}

Evaluating the expression at $r=d$ gives the speed of sound in the thin shell. According to Poisson and Visser, causality requires the speed of sound not to exceed the speed of light $(c=1)$, implying $0 \leq \eta \leq 1$.

Figure \ref{10} shows that the causality condition $0 \leq \eta \leq 1$ constrains the shell thickness. The stability of gravastars has been extensively studied using bounds on the speed of sound. However, analyzing stability via speed of sound has limitations. Specifically, for stiff matter $(\omega = 1)$, the expression may not accurately represent the speed of sound due to unknown microscopic degrees of freedom.
This stability condition provides a necessary condition for the stability of the thin shell gravastar model. From the figure \ref{10} it is clear that our proposed model in this gravity lies with the stability range that is $0$ and $1$ in this gravitational framework which leads to a viable gravastar model.

\subsection{Surface redshift}
\hspace{0.5cm}The surface redshift of gravasters is crucial for understanding stability and detection. Defined as $Z_s =\frac{\Delta \lambda}{\lambda_{e}}=\frac{\lambda_{0}}{\lambda_{e}}$, it measures the fractional change in wavelength. For isotropic static perfect fluids, $Z_{s}<2$, while anisotropic fluids can have $Z_{s}$ up to 3.84 \cite{Buchdahl,Ivanov}. Surface redshift limit vary: for isotropic fluids without cosmological constant, $Z_{s}\leq2$ \cite{Barraco2002}, while for anisotropic stars with cosmological stars with cosmological constant, $Z_{s}$ can be $\leq5$ \cite{Böhmer2006}. We thus calculate the surface redshift using 
\begin{equation}
  Z_{s}=-1+\mid g_{tt}\mid^{-\frac{1}{2}}.
\end{equation}
The surface redshift is ultimately obtained as

\begin{equation}
 Z_{s}=-1+\frac{e^{\frac{-Br^2}{2}}}{C} 
\end{equation}

\begin{figure}[h!]
  \centering
  \includegraphics[width=0.4\textwidth]{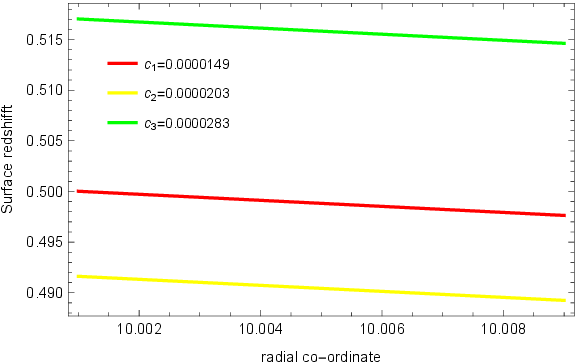}
  \caption{The surface redshift of the shell changes with its thickness, measured in $r$ for $C_{1}=0.0000149, C_{2}=0.0000203, C_{3}=0.0000283$}
   \label{11}
\end{figure}

Figure \ref{11} shows the variation of surface redshift, which remains within the stability range throughout the thin shell, indicating that our gravastar model is stable and physically acceptable.

\section{Energy conditions}
\hspace{0.5cm}Using the  equations below energy conditions are defined as:\\
1. Weak Energy Condition (WEC): $\sigma \geq 0$ and $\sigma+\upsilon\geq 0$.\\
2. Null Energy Condition (NEC):  $\sigma+\upsilon\geq 0$.\\
3. Strong Energy Condition (SEC): $\sigma+\upsilon\geq 0$ and  $\sigma+3\upsilon\geq 0$.\\
4. Dominant Energy Condition (DEC): $\sigma\geq 0$, $\sigma+\upsilon\geq 0$ and  $\sigma-\upsilon\geq 0$.\\
These conditions help to determine the physical viability of spacetime solutions. The plots show WEC, NEC and SEC are satisfied for the thin shell, but DEC is violated, implying all observers would measure a negative energy flux.
\begin{figure}[h!]
  \centering
  \includegraphics[width=0.45\textwidth]{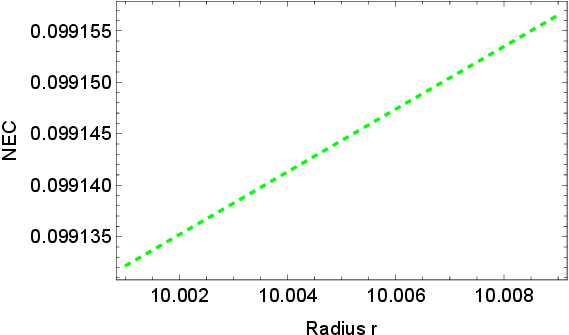}
  \caption{NEC condition plot with respect to r for $M = 0.338m_{\bigodot}, A=-2.5,\rho_{0}=0.01,\aleph=10$.}
   \label{12}
\end{figure}
\begin{figure}[h!]
  \centering
  \begin{minipage}[b]{0.6\textwidth}
    \centering
    \includegraphics[width=0.75\textwidth]{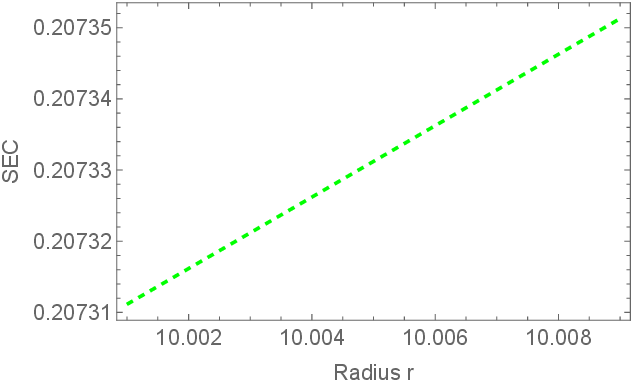}
    \caption{SEC condition plot with respect to r for $M = 0.338m_{\bigodot}, A=-2.5,\rho_{0}=0.01,\aleph=10$.} 
    \label{13}
  \end{minipage}
  \hfill
  \begin{minipage}[b]{0.39\textwidth}
    \centering
    \includegraphics[width=1.1\textwidth]{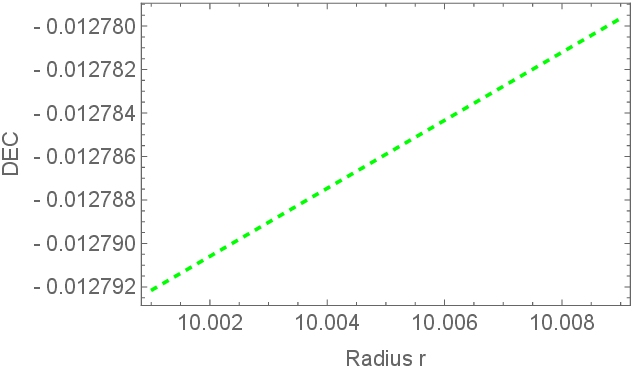}
    \caption{DEC condition plot with respect to r for $M = 0.338m_{\bigodot}, A=-2.5,\rho_{0}=0.01,\aleph=10$.} 
    \label{14}
  \end{minipage}
\end{figure}

\section{Discussion and conclusion}

\hspace{0.5cm}This paper presents a new class of gravastar solutions using the Kuchowicz metric potential within the framework of $f(R, \Sigma, T)$ gravity. The Kuchowicz metric potential is adopted due to its singularity-free nature, ensuring regular behavior throughout the gravaster's structure. We've explored the characteristics of the gravastar's three regions (interior, shell, and exterior) with the help of Kuchowicz metric potential. The solutions for the interior region of the gravastar are regular and well-behaved at the center $(r=0)$, avoiding any singularities. The calculated properties of the shell demonstrate physical viability and acceptability of the gravastar model in this study. The gravaster model, proposed by Mazur and Mottola, has garnered significant attention and research, indicating its potential to address issues associated with black holes, such as singularities and information paradoxes. This section will summarize key findings and important results derived from the current study on gravastars.\\ 

\noindent\textbf{(1) Interior region:} By solving the system of eqs. \eqref{50}–\eqref{52} along with the conservation eq. \eqref{48}, we obtain solutions for the metric function, pressure, and matter density within the interior region of the gravastar. The conservation equation reveals that both pressure and matter density are constant throughout the interior region of the gravastar. The central density of the gravaster can be linked to Einstein's vacuum energy density in the presence of a cosmological constant, providing an interesting connection between the two concepts. The constant matter density provides an outward pressure that contributes to the stability of the gravaster. The metric potential's variation with radial distance is also illustrated in Fig. \ref{1}, showcasing its behavior. The interior region's active gravitational mass is singularity-free, being zero at the origin and positive throughout the interior.\\ 
\noindent\textbf{(2) The intermediate thin shell:} 
The thin shell region of the gravastar is composed of an ultra-relativstic fluid, specifically a stiff fluid model, characterized by high-density soft quanta, as described by Zel's dovich \cite{Zel’dovich}. The field eqs. \eqref{50}–\eqref{52} and conservation eq. \eqref{48} were solved to determine the metric function and characteristics of the shell relying on the thin shell approximation. A set of exact solutions for the shell's properties has been obtained, providing a description of its characteristics.\\  
\noindent\textbf{(3) Exterior region:} The exterior region of the gravastar is a vacuum $(P=\rho=0)$ and is described by the Schwarzschild metric, as give in eq. \eqref{68}.\\
\noindent\textbf{(4) Certain physical properties of the shell:} Despite  the shell's small but finite thickness, key physical parameter have been investigated, proper thickness, energy and entropy. The behavior of these parameters across the shell is illustrated in Figs. \eqref{4}-\eqref{6}. The energy and entropy of the shell follow a similar trend, likely decreasing from the interior to exterior boundary, consistent with the behavior of the density. Figures \eqref{6} and \eqref{5} illustrate the variation of energy and entropy, respectively, and the expected trend. Fig. \eqref{4} depicts the variation of proper thickness. These figure collectively demonstrate the physical viability and consistency of the model, supporting its validity. \\ 
\noindent\textbf{(5) Junction conditions:} Junction conditions were carefully examined to create a thin shell connecting the inner and outer spacetimes. Figures \eqref{7} and \eqref{8} illustrate the relationship between surface energy density and surface pressure with respect to $r$. The interior solution has been matched to the exterior Schwarzschild metric at the boundary, with the thin shell in between, ensuring a smooth transition. The discontinuity is the second fundamental from at the junction interface suggests the presence of a shell filled with ultra-relativistic fluid, characterized by an EoS $P=\rho$, along with matter components. The two fluids, one ultrarelativistic ($P=\rho$) and the other component, are non-interacting and together characterize the shell of the gravastar.

\noindent\textbf{(6) Stability checking:} The surface redshift and speed of sound for the shell has been calculated, following \cite{Banerjee2018}, to investigate the stability of the gravastar model. The surface redshift and the sped of sound for the model is found to be within the stability range, indicating a comprehensive approach to validate the model's robustness. The stability of a lower-dimensional gravastar was explored using perturbative treatment and linearized stability analysis for the shell dynamics \cite{Banerjee2016}. Other stability checks, such as dynamical stability against radial perturbations or axial perturbations, are also possible but were not conducted in this study \cite{DeBenedictis2006}. Given the physically acceptable results of our model, it is plausible that model may also exhibit dynamic stability against both radial and axial perturbations, although further analysis would be needed to confirm this.
Figs. \ref{10} and \ref{11} specifically showing how these stability parameters varies with radius. The figures show that both remains positive and within 1, confirming the stability of our proposed model. \\
\noindent\textbf{(7)Energy conditions} The energy conditions WEC, NEC, and SEC are satisfied for the thin shell's but the Dominant Energy Condition (DEC) is violated for the thin shell construction, as shown in figures \ref{12}-\ref{14}.\\
This study explored a gravaster model using a distinct approach, specifically utilizing the Kuchowicz  metric potential, offering a fresh perspective on this gravitational theory. By fixing one parameter, the search for physically acceptable solutions for the remaining parameters becomes more constrained and potentially more challenging. The study yields physically acceptable  solution that remain finite at the origin, suggesting that the metric potential provides a theoretically sound and viable framework for constructing gravastars. These findings encourage further research into gravastars using various metric potentials. Although gravastars haven't been directly observed, the detection of gravitational waves by LIGO sparks debate about whether the waves originate from black hole or gravastar mergers, warranting continued exploration. This article demonstrates the theoretical possibility of gravastars, showcasing their physical viability and potential existence.

\end{document}